\pdfoutput=1
\documentclass[a4paper,11pt]{article}
\usepackage[dvipsnames]{xcolor}
\usepackage{amsmath,amssymb,amsthm,amsfonts,mathrsfs}
\usepackage{bm}
\usepackage{booktabs, comment}
\usepackage{tabularx}
\usepackage{jheppub}
\usepackage[T1]{fontenc}
\usepackage[utf8]{inputenc}
\usepackage{hyperref}
\hypersetup{
	linktoc=all,
	colorlinks=true,
	linkcolor=blue,
	citecolor=magenta,
	urlcolor=red
}

\usepackage[most]{tcolorbox}

\tcbset{highlight math style={left=02mm,right=02mm,top=02mm,bottom=02mm}} 
\usepackage{empheq}

\newcommand\inbox[1]{\tcbset{fonttitle=\scriptsize} \tcboxmath[colback=white,colframe=black!70]{#1}}

\newcommand{\cV}{\mathcal{V}}
\newcommand{\cW}{\mathcal{W}}

\newcommand{\dd}{\mathrm{d}}
\makeatletter \@addtoreset{equation}{section}

\newcommand\bc{\begin{center}}
	\newcommand\ec{\end{center}}
\begin{document}
	\vspace*{-5cm}
	\title{\bc
		\huge{What is Classical Null String Theory? }
		\ec}
	
	\author[a]{M.M. Sheikh-Jabbari }
	\author[b]{and H. Yavartanoo}
	
	\affiliation[a]{School of Physics, Institute for Research in Fundamental Sciences (IPM),
		P.O. Box 19395-5531, Tehran, Iran}
	\affiliation[b]{Beijing Institute of Mathematical Sciences and Applications (BIMSA),
		Huairou District, Beijing 101408, P. R. China}
	
	\emailAdd{jabbari@theory.ipm.ac.ir}
	\emailAdd{yavar@bimsa.cn}

\abstract{We revisit the classical definition of a null string in a
$D$-dimensional Minkowski target space and distinguish a consistent
partially-gauged Carrollian $\sigma$-model from the null string theory. The 
Isberg--Lindstr\"om--Sundborg--Theodoridis (ILST) action gauges
worldsheet diffeomorphisms and the common Weyl rescaling, but not the
independent relative rescaling of the temporal and spatial
representatives of the intrinsic Carrollian structure, Carroll-Weyl scaling. A physical string theory  requires every local change of worldsheet representative to be gauged, therefore, the  Carroll--Weyl symmetry should be gauged in a null string theory. In the minimal realization, the corresponding gauging is obstructed at finite tension, whereas at zero tension it yields an additional first-class constraint and the associated identification of physical configurations along its gauge orbits. The description of the null worldsheet as a congruence of null geodesics provides a complementary target-space interpretation of gauging of the Carroll-Weyl scaling. Our analysis thus provides the supplementary requirements that turns the ILST theory into the null string theory.} 
	\maketitle
\section{Introduction}\label{sec:introduction}
String theory is about studying the dynamics of one-dimensional objects, strings, sweeping out $1+1$ dimensional worldsheets while probing a $D$-dimensional target space. 
The standard tensile string admits two classically equivalent formulations: the reparametrization-invariant Nambu--Goto action and the Polyakov action \cite{Polyakov:1981rd}, with an independent worldsheet metric \cite{Green:1987sp,Polchinski:1998rr}. The former yields the picture of a tensile string as an extremal-area surface in the target space, subject to the usual constraints. The Polyakov action, on the other hand, focuses more directly on the worldsheet and its symmetries, asserting that the local redundancies of the Lorentzian worldsheet description, namely two-dimensional ($2d$) diffeomorphisms and Weyl scaling, should be treated as gauge symmetries. If two configurations differ only by a local change of representative of the same intrinsic worldsheet
geometry, but the theory treats them as physically distinct, then the theory overcounts its physical
configurations. Therefore, gauging all such worldsheet redundancies is a necessary, although not sufficient, condition for the $\sigma$-model to describe a physical string. This logic applies to tensile as well as to null string theory.

In a $d+1$ dimensional Lorentzian geometry, one can consider $d$-dimensional null hypersurfaces such as cosmological and black hole horizons or null infinity in asymptotically flat spacetimes. The induced geometry on these null hypersurfaces is called Carrollian \cite{Bacry:1968zf, Duval:2014uoa, Duval:2014lpa,  Hartong:2015xda, Ciambelli:2019lap, Herfray:2021qmp, Ciambelli:2025unn}, which have a degenerate metric with a  vanishing eigenvalue. Carrollian geometries may be studied intrinsically, without reference to an embedding. Alternatively, one may obtain a $d$-dimensional Carrollian geometry through an ultra-relativistic contraction of a $d$-dimensional Lorentzian geometry, in which the light cones pinch off. Since this is a singular limit, it may obscure intrinsic and essential aspects of the (gauge) symmetries of the resulting Carrollian geometry. An independent intrinsic formulation of Carrollian geometry therefore provides the appropriate framework for identifying its geometric redundancies.

Null strings, namely strings with Carrollian worldsheets, have been studied since 1976 \cite{Schild:1976vq}. They are usually viewed as the tensionless limit of tensile strings, and their action can be obtained from the Polyakov action in the vanishing-tension limit through the Isberg--Lindstr\"om--Sundborg--Theodoridis (ILST) formulation \cite{Isberg:1992ia,Isberg:1993av}. See \cite{Bagchi:2026wcu} and references therein for a recent review of null strings. While one may obtain a new theory through a singular limit of a given theory, applying the same limit to its symmetry analysis and gauge reduction requires special care. The symmetry algebra may undergo an In\"on\"u--Wigner contraction, and additional gauge redundancies may emerge in the limiting geometry. 
The main thrust of this work is establish that in the ILST action, obtained as the tensionless limit of the tensile string,  all the Carrollian worldsheet redundancies do not appear as gauge symmetries. Therefore, while the ILST action describes a consistent theory, it does not by itself describe a theory of physical null strings.

In this paper, we tackle the basic question that is logically prior to quantization: \emph{what intrinsic data define the null string?} We answer this question implementing the criterion outlined in the first paragraph of the introduction. Carrollian worldsheets, besides the $2d$ diffeomorphisms and Weyl scaling, also enjoy Carroll-Weyl scaling, and the Worldsheets related by these transformations are physically identical. So, null string theory should have the three symmetries as gauge symmetries. In this work, we construct a \textit{minimal extension} of the ILST action which realizes Carroll-Weyl scaling as gauge symmetry along with the $2d$ diffeomorphisms and Weyl scaling. We also provide a target-space interpretation of the gauged action and the extra constraint and identifications that the ILST action should be supplemented with  to describe a physical null string theory.
\section{Null strings are described by the Carroll-Weyl gauged action}
\label{sec:NS-gauged-action}

The central distinction is most cleanly formulated from the \textit{worldsheet} viewpoint. We first recall the difference between a consistent $\sigma$-model and a physical string theory, and then apply the same criterion to the intrinsic Carrollian geometry of the null worldsheet.

\subsection{A quick review of ILST action}\label{sec:ILST-review}

Traditionally, it is said that ``null string theory'' on flat $D$ dimensional Minkowski background is described by the ILST action \cite{Isberg:1993av}
\begin{equation}
 S_{\rm ILST}
 =\frac{\kappa}{2}\int_{\Sigma}d^2\sigma\,
 \mathcal V^a\mathcal V^b
 \eta_{\mu\nu}
 \partial_aX^\mu\partial_bX^\nu ,
 \label{eq:ILST-action-physical}
\end{equation}
where $\sigma^a=(\tau,\sigma)$ parametrizes the $2d$ Carrollian worldsheet
and $\mathcal V^a$ is a contravariant worldsheet vector density of weight
$1/2$. The Carrollian worldsheet is described by a kernel vector along
$\mathcal V^a$ and a degenerate metric $h_{ab}$ with one vanishing
eigenvalue. The metric $h_{ab}$ does not explicitly appear  in
\eqref{eq:ILST-action-physical}; see \cite{Sheikh-Jabbari:2026vqh} and
Appendix~\ref{app:geometry} for more details. The action is invariant under
$2d$ diffeomorphisms and the standard Weyl scaling, which are treated as
gauge symmetries, as in the Polyakov action before taking the Carrollian
limit \cite{Isberg:1993av,Bagchi:2026wcu,Sheikh-Jabbari:2026vqh}. 

One may partially fix the gauge symmetries by choosing the temporal gauge
\begin{equation}
 \mathcal V^a=(1,0).
 \label{eq:ILST-temporal-gauge}
\end{equation}
The equations of motion for the embedding coordinates $X^\mu$ and the
components of $\mathcal V^a$ then become
\begin{equation}\label{eq:ILST-temporal-equations}
 \begin{split}
  \ddot X^\mu:=&\dot P^\mu=0,\\
  P^2=0,
  \qquad&
  P\cdot X'=0,
 \end{split}
\end{equation}
where $\dot X:=\partial_\tau X$ and $X':=\partial_\sigma X$. These equations
are solved by
\begin{equation}
 X^\mu(\tau,\sigma)
 =P^\mu(\sigma)\tau+Q^\mu(\sigma),
 \label{eq:ILST-general-solution}
\end{equation}
subject to
\begin{equation}
 C_1=P^2(\sigma)=0,
 \qquad
 C_2=P\cdot Q'(\sigma)=0,
 \label{eq:ILST-two-constraints}
\end{equation}
and the identification of configurations related by the residual
diffeomorphism and standard Weyl transformations
\cite{Sheikh-Jabbari:2026cnj,Sheikh-Jabbari:2026tpf}.

\subsection{\texorpdfstring{$\sigma$-model}{sigma-model} vs. string theory \& Lorentzian vs. Carrollian}\label{sec:physical-motivation}

The ILST theory as briefly reviewed above,  is a consistent action in its own right; the question is which physical system it describes. To answer, it is helpful to note the distinction between a $\sigma$-model and a string theory. The degrees of freedom of a $\sigma$-model are maps from a fixed $2d$ background into a $D$-dimensional target space. A string theory is not specified only by the equations of motion of its embedding fields. One must also decide which changes of the intrinsic worldsheet data alter the geometry and which merely change its representative. Every local change of representative must be gauged. This is a necessary, although not sufficient, condition for a sigma model on a $2d$ background to describe physical strings. For the tensile string this statement is familiar. Any two worldsheet metrics related by a diffeomorphism and/or a Weyl transformation represent the same worldsheet and the same string. A sigma model on a fixed Lorentzian metric may be perfectly consistent, but that fact alone does not make it a string theory. Thus, the Polyakov action, in which $2d$ diffeomorphisms and  Weyl rescalings \cite{Polyakov:1981rd} are gauged, describes a string theory. After gauge fixing which completely fixes the $2d$ metric, one should still impose the Virasoro constraints as the residual gauge symmetries \cite{Green:1987sp,Polchinski:1998rr}.

By the same token, one must distinguish a Carrollian $\sigma$-model, whose degrees of freedom are maps from a fixed $2d$ Carrollian background into a target space, from a null-string theory with fully dynamical intrinsic Carrollian data and all their local representative redundancies gauged. Applying the controlled ultra-relativistic contraction to the Polyakov action yields the ILST action \cite{Isberg:1993av}. The nondegenerate Lorentzian worldsheet geometry under the limit becomes a degenerate Carrollian structure, and importantly, the local changes of representative of the limiting geometry are no longer exhausted by simply limiting the $2d$ diffeomorphisms and Weyl scaling. The ILST action is therefore a consistent, partially gauged, Carrollian $\sigma$-model; but by itself it is not a  null-string theory (in which all null worldsheet symmetries are gauged).

As pointed out in \cite{Sheikh-Jabbari:2026vqh} and reviewed below, a $2d$ Carrollian geometry admits, besides diffeomorphisms, two independent local Weyl rescalings. One is the common rescaling inherited from the tensile worldsheet; the other, which we call Carroll-Weyl scaling, changes the relative normalization of the temporal and spatial Carrollian directions while leaving the underlying line fields unchanged. Both are changes of representative of the same intrinsic Carrollian geometry. Therefore, both must be gauged in a physical null-string theory. {The standard ILST limit retains the common Weyl redundancy but not the  relative Carroll--Weyl scaling;  it exhibits a restricted remnant of the Carroll-Weyl gauge symmetry, the $\chi$-symmetry of \cite{Sheikh-Jabbari:2026cnj}. Thus, the Gauss-law  constraint associated with Carroll-Weyl gauge symmetry, while is admissible in the ILST theory, is not necessitated by it. Consequently, the ILST action can be consistently promoted to a null string theory once supplemented with the constraint resulting from the Carroll-Weyl gauging and quotienting the string configurations by the orbits of the gauge symmetry.}

The above discussion may be formulated as in
\cite{Sheikh-Jabbari:2026vqh}, and here we briefly review it for
completeness. A $2d$ Carrollian worldsheet is locally described through 
$(n_a,\ell_a;v^a,\ell^a)$, satisfying
\begin{equation}\label{eq:two-Carroll-Weyl}
 v^an_a=1,
 \quad v^a\ell_a=0,
 \quad \ell^an_a=0,
 \quad \ell^a\ell_a=1.
\end{equation}
The degenerate metric $h_{ab}=\ell_a\ell_b$ and volume form
$\varepsilon_{ab}=n_a\ell_b-n_b\ell_a$ are derived from these same data.
The two rescalings act on the coframe and dual frame together,
\begin{equation}
\begin{split}
 n_a\longmapsto e^{\chi_t}n_a,&\qquad
 v^a\longmapsto e^{-\chi_t}v^a,
 \\
 \ell_a\longmapsto e^{\chi_s}\ell_a,&\qquad
 \ell^a\longmapsto e^{-\chi_s}\ell^a,
\end{split}
\end{equation}
with two independent functions $\chi_t,\chi_s$.
Thus, the discussion concerns different representatives of one complete
Carrollian structure, not alternative choices between line fields and a
degenerate metric. Consequently, both rescalings are redundancies of the
worldsheet description and must be gauged. Explicitly, physical worldsheets are defined up to 
$\big(n_a,\ell_a;v^a,\ell^a\big)\sim \big(e^{\chi_t} n_a,e^{\chi_s}\ell_a;e^{-\chi_t}v^a,e^{-\chi_s}\ell^a\big)$ equivalence classes.

The $\chi_t=\chi_s$ sector is a gauge symmetry of the ILST action, since
${\cal V}^a$ remains invariant under this transformation
\cite{Sheikh-Jabbari:2026vqh}; this sector is inherited from the Weyl
symmetry of the tensile string before taking the limit. The
$\chi_t=-\chi_s:=\chi$ sector emerges for the null string. Under this
transformation, as shown in Appendix~\ref{app:geometry},
\begin{equation}
 \cV^a\longmapsto e^{-\chi}\cV^a.
 \label{eq:V-relative-weight}
\end{equation}
With inert target coordinates $X^\mu$, the ILST action
\eqref{eq:ILST-action-physical} is therefore a consistent partially gauged Carrollian
$\sigma$-model, but it does not by itself describe a physical null string:
it gauges the $2d$ diffeomorphisms and the common Weyl rescaling, while the
relative Carroll--Weyl redundancy \eqref{eq:V-relative-weight} remains
ungauged.

\subsection{The gauged action and the third constraint in Minkowski space}
\label{sec:CW-gauging}

To state the main point, consider the ILST action in \eqref{eq:ILST-action-physical}. To turn the ILST action into a
physical null string theory, we must realize the $\chi$ transformation
\eqref{eq:V-relative-weight} as a gauge symmetry. This can be \textit{minimally} (without introducing any other worldsheet or target-space field and keeping it a two-derivative theory) achieved  by
assigning a Carroll--Weyl weight to $X^\mu$ and replacing the worldsheet
derivatives with a Carroll--Weyl covariant derivative
\cite{Gustafsson:1994kr,Sheikh-Jabbari:2026vqh,Lindstrom:2026zno}.
Explicitly, the gauged action is
\begin{equation}
 S_{\rm CW}
 =\frac{\kappa}{2}\int_\Sigma\dd^2\sigma\,
 \cV^a\cV^b\eta_{\mu\nu}D_aX^\mu D_bX^\nu,
 \label{eq:main-CW-action}
\end{equation}
with
\begin{equation}
 D_aX^\mu=\partial_aX^\mu+\cW_aX^\mu,
 \label{eq:main-CW-derivative}
\end{equation}
and is invariant under the Carroll--Weyl gauge transformations
\begin{equation}\label{eq:main-diagonal-CW}
 \delta_\chi\cW_a=-\partial_a\chi,
 \qquad
 \delta_\chi X^\mu=\chi X^\mu,
 \qquad
 \delta_\chi\cV^a=-\chi\cV^a.
\end{equation}
Here, $\cW_a$ is the Carroll--Weyl connection. We note that in the finite form the above takes the form, $X^\mu\mapsto e^\chi X^\mu, \cV^a\mapsto e^{-\chi} \cV^a$ \cite{Sheikh-Jabbari:2026vqh}.

A full classical analysis of this action may be found in
\cite{Sheikh-Jabbari:2026vqh,Sheikh-Jabbari:2026tpf}. Here, we only mention
that the equation of motion for $\cW_a$ yields the Carroll--Weyl Gauss-law
constraint
\begin{equation}\label{C3-const}
 P\cdot X=0,
\end{equation}
which on shell reduces to $C_3:=P\cdot Q=0$. After imposing the Gauss-law
constraint, one may choose the $\cV^a\cW_a=0$ gauge. It is important that the Carroll--Weyl Gauss-law constraint is obtained and imposed together with the gauge-fixing. Gauge-fixing the action before varying $\cW_a$ would incorrectly discard this constraint. The gauge condition leaves residual transformations satisfying
\begin{equation}
 \cV^a\partial_a\chi=0,
\end{equation}
which in the temporal gauge \eqref{eq:ILST-temporal-gauge} becomes
$\chi=\chi(\sigma)$. As discussed in
\cite{Sheikh-Jabbari:2026cnj}, the ILST action exhibits precisely this
restricted $\chi(\sigma)$ symmetry. In this gauge, the gauged action reproduces the ILST action
\eqref{eq:ILST-action-physical}, supplemented by the Carroll--Weyl
Gauss-law constraint \eqref{C3-const} and $X^\mu\sim e^{\chi(\sigma)} X^\mu$ identifications. 

{Consideration of the $\chi$-symmetry yields an extension of BMS$_3$ algebra, which arises after the temporal gauge-fixing \eqref{eq:ILST-temporal-gauge}. This extended BMS$_3$ plays the same role for null string theory as the Virasoro algebra for tensile strings: Null strings are states/configurations of the ILST action subject to this extended BMS$_3$ algebra \cite{Sheikh-Jabbari:2026cnj, Rasulian:2026jvg}. } 

We should emphasize that the existence of this restricted $\chi$-symmetry does not, by itself, explain why
its conserved charge should vanish or why the transformation should be
gauged. Indeed, the ILST action admits a larger class of restricted
symmetries associated with target-space (flat $D$ dimensional Minkowski spacetime) conformal Killing vectors
$K^\mu$, with corresponding conserved quantities
$C_K(\sigma)=P\cdot K$. These charges should not in general be set to
zero. The case $K^\mu=X^\mu$ is singled out here for a different reason:
{For general $K$ we are not dealing with gauging the worldsheet Carroll-Weyl scaling. See appendix \ref{app:gauging-CKV} for more analysis. Explicitly, the gauged action \eqref{eq:main-CW-action} is  \textit{the minimal} action which realizes $2d$ diffeomorphisms, conventional Weyl and Carroll-Weyl gauge symmetries, which yield $P\cdot X=0$ physicality constraint. }

We close this section by a comment on how the tensile-string constraints transform under the local scaling generated by $P\cdot X$. For the tensile string, the constraints are
\begin{equation}
    C^{\text{T}}_1=\frac12 (P^2+T^2 X'{}^2),\qquad C^{\text{T}}_2=P\cdot X',
\end{equation}
where $T$ is the string tension, transform as
\begin{equation}\label{delta-C-T}
\delta_\chi C^{\text{T}}_1\simeq T^2\left[\chi'X\cdot X'+2\chi X'{}^2\right],\qquad \delta_\chi C^{\text{T}}_2\simeq \chi' P\cdot X\,.
\end{equation}
The obstruction therefore disappears in the tensionless limit, making it possible for the tensionless string to
acquire the additional gauge symmetry.

The logical order is important. The disappearance of the tensile
obstruction shows that $P\cdot X$ can consistently be treated as a
first-class constraint at zero tension, but it does not by itself explain
why the constraint must be imposed. Its necessity follows from the
intrinsic Carrollian geometry and the requirement that every local change
of worldsheet representative be gauged. Its admissibility follows from the
zero-tension constraint algebra. Thus, the additional gauging is not an
optional enlargement of the physical null string theory; it is required
for a gauge-complete description of the null worldsheet.

\section{Null strings as a congruence of null geodesics}
\label{sec:null-string-target-space}

Having established the necessity of Carroll--Weyl gauging from the intrinsic worldsheet geometry, we now describe how this gauging appear and act in a more intuitive target-space description. We restrict the discussion to the ILST action on a
$D$-dimensional flat Minkowski background \eqref{eq:ILST-action-physical} in the temporal gauge \eqref{eq:ILST-temporal-gauge} and classical solution \eqref{eq:ILST-general-solution} subject to constraints \eqref{eq:ILST-two-constraints}. We restrict attention
to configurations with $P^\mu(\sigma)\neq0$.

Eq.~\eqref{eq:ILST-general-solution} specifies a one-parameter family (congruence) of affinely parametrized null geodesics. The coordinate
$\tau$ is an affine parameter along each null generator, while $\sigma$
labels the generators whose union forms the ruled null worldsheet. The
shape of the congruence is specified by $Q^\mu(\sigma)$. The constraint
$C_1=0$ implies that $P^\mu(\sigma)$ is null, while $C_2=0$ implies that
$Q'^\mu(\sigma)$ lies in the hyperplane orthogonal to $P^\mu(\sigma)$.

An affinely parametrized null geodesic has no preferred normalization of its
affine parameter. One may rescale the parameter by a nonzero constant and
shift its origin without changing the underlying null geodesic. Along a
congruence, these freedoms may depend on $\sigma$. It is important,
however, to distinguish this affine reparametrization from the additional
Carroll--Weyl redundancy. Consider first a pure change of affine parameter,
\begin{equation}
 \widetilde\tau=a(\sigma)\tau+b(\sigma).
 \label{eq:pure-affine-transformation}
\end{equation}
We restrict to $a(\sigma)>0$. \footnote{This restriction preserves the
orientation of the worldsheet time direction.} {This is a worldsheet diffeomorphism under which $\cV^a\to a(\sigma)^{\frac12} \cV^a$ but does not involve $\chi$-scaling. Under this diffeomorphism, $\tilde X^\mu (\tilde \tau, \sigma)= X^\mu(\tau, \sigma)$, yielding,}
\begin{equation}
 \widetilde P^\mu=a^{-1}P^\mu,
 \qquad
 \widetilde Q^\mu
 =Q^\mu-\frac{b}{a}P^\mu.
 \label{eq:pure-affine-PQ}
\end{equation}
Consequently,
\begin{equation}
 \widetilde C_1=a^{-2}C_1,
 \qquad
 \widetilde C_2
 =a^{-1}\left[
 C_2-\left(\frac{b}{a}\right)'C_1
 -\frac{b}{2a}C_1'
 \right].
 \label{eq:pure-affine-constraints}
\end{equation}
Therefore, {as expected}, a pure affine reparametrization preserves the constraint surface
$C_1=C_2=0$ without requiring an additional constraint. In particular,
$P\cdot Q=0$ does not follow merely from the freedom to change the affine
parameter. {Nonetheless, it is consistent with the diffeomorphism, as $\tilde P\cdot \tilde Q= a^{-1} P\cdot Q- \frac{b}{a^2}C_1$, so that upon $C_1$, $\tilde P\cdot \tilde Q=0$ follows from $P\cdot Q=0$.}

The third constraint instead arises from the additional Carroll--Weyl
change of worldsheet representative. {To see, let us supplement the diffeomorphism in \eqref{eq:pure-affine-transformation} with a $\chi$-transformation to keep $\cV^a$ intact. This may be achieved by
\begin{equation}
 \widetilde\tau=\alpha(\sigma)\tau+\beta(\sigma),\qquad \chi=\frac12\ln \alpha(\sigma), \qquad \alpha(\sigma)>0.
 \label{eq:chi-affine-transformation}
\end{equation}
Under this transformation $\tilde X^\mu (\tilde\tau, \sigma)= \sqrt{\alpha}\ X^\mu (\tau, \sigma)$
\begin{subequations}\label{eq:residual-PQ}
 \begin{align}
  \widetilde P^\mu&=\frac1{\sqrt{\alpha}} P^\mu,
  &
  \widetilde Q^\mu&=\sqrt{\alpha} Q^\mu-\frac{\beta}{\sqrt{\alpha}} P^\mu,
  \label{eq:residual-PQ-fields}\\
  \widetilde C_1&=\frac1\alpha C_1,
  &
  \widetilde C_2&=C_2-\frac{\beta}{2\alpha} C_1'- \frac{\beta}{\alpha}(\frac{\beta'}{\beta}-\frac{\alpha'}{2\alpha}) C_1 +\frac{\alpha'}{2\alpha} P\cdot Q.
  \label{eq:residual-PQ-constraints}
 \end{align}
\end{subequations}
Here, $\alpha(\sigma)$ and $\beta(\sigma)$ are arbitrary residual gauge
parameters.} While $C_1=0$ implies $\widetilde C_1=0$, the conditions $C_1=C_2=0$ do not by themselves imply $\widetilde C_2=0$. Since
$\alpha(\sigma)$ is an arbitrary Carroll--Weyl gauge parameter,
preservation of the constraint surface under
\eqref{eq:residual-PQ} requires
\begin{equation}
 C_3(\sigma):=P\cdot Q=0.
 \label{eq:third-constraint-PQ}
\end{equation}
This is the temporal-gauge form of the Carroll--Weyl Gauss-law constraint
$P\cdot X=0$, because
\begin{equation}
 P\cdot X=\tau P^2+P\cdot Q.
\end{equation}
{The transformations \eqref{eq:chi-affine-transformation} recover Carroll-Weyl scaling and $\tau$-reparametrizations in the temporal gauge, whereas the null string theory also involves the $\sigma$-reparametrizations, $\sigma\to \tilde\sigma=\gamma(\sigma)$. See \cite{Sheikh-Jabbari:2026cnj} for the infinitesimal form of these transformations. $P\cdot Q'=0$ is the first-class constraint associated with $\sigma$-reparametrizations, while $P^2=0$ is the one corresponding to $\beta(\sigma)$ and $P\cdot Q=0$ the one corresponding to $\alpha(\sigma)$ \cite{Sheikh-Jabbari:2026tpf}.} The gauge-complete null string is therefore described by a congruence of null geodesics satisfying
\begin{equation}\label{Null-string-def}
 \inbox{
 P^2=0,\qquad
 P\cdot Q'=0,\qquad
 P\cdot Q=0,\qquad
 (P,Q)\sim
 \bigl( \frac1{\sqrt{\alpha}} P,\sqrt{\alpha} Q-\frac{\beta}{\sqrt{\alpha}} P\bigr)
 }
\end{equation}
together with identifications under the residual $\sigma$-reparametrizations. The first two
conditions are the standard ILST constraints, whereas the third condition
and its associated identification implement the additional
Carroll--Weyl gauge redundancy.

Geometrically, $P^2=0$ makes each generator null, while $P\cdot Q=0$
implies $Q^\mu(\sigma)$  lies in the hyperplane orthogonal to $P^\mu(\sigma)$. In a generic congruence,
however, $P^\mu(\sigma)$ varies with $\sigma$, so these hyperplanes need
not combine into one fixed codimension-one hypersurface in the Minkowski
target space. The general target-space statement is instead that the third
constraint generates a local scaling identification and removes one
additional canonical pair. Together with the two standard ILST constraints
and their gauge symmetries, this leaves $D-3$ local configuration-space
degrees of freedom on the regular part of the constraint surface
\cite{Sheikh-Jabbari:2026tpf}.

A more detailed geometric interpretation of this target-space
identification lies beyond the scope of the present work and will be presented in an upcoming publication.

\section{Concluding remarks}
\label{sec:conclusion}

The ILST action correctly describes the motion of a degenerate worldsheet
and is a consistent Carrollian \textit{partially gauged $\sigma$-model}.
This is necessary, but not sufficient, to identify the theory of a physical
null string. A physical string theory must be defined on equivalence classes
of intrinsic worldsheet data, so every local transformation that changes
only the representative of that geometry must be gauged. In doing so, one
should retain the minimal field content on the worldsheet and in the target
space. In a partially gauge-fixed form, \textit{the ILST action describes a
physical null string only when supplemented with the Carroll--Weyl
constraint $P\cdot Q=0$ and the gauge quotient it generates.}

The ILST action remains a classically consistent Carrollian partially-gauged 
$\sigma$-model; the distinction concerns its gauge completeness as a
physical null string theory. In temporal gauge, its restricted $\chi(\sigma)$ symmetry
implies that the solutions are organized into sectors labelled by the
conserved quantity $C_3(\sigma)=P\cdot Q$. The existence of this symmetry
does not require its charge to vanish. In the fully gauged theory, however,
the equation of motion of the Carroll--Weyl connection imposes
$C_3(\sigma)=0$ as a Gauss-law constraint, and configurations along the
corresponding gauge orbits must be identified. Choosing
$\cV^a\cW_a=0$ then reproduces the ILST form of the action, but only after
the Gauss-law constraint has been derived and retained. Thus, the
gauge-fixed Lagrangian may coincide with the ILST Lagrangian while the two
theories still have different physical phase spaces.

The third constraint $C_3$ is independent of the two standard ILST constraints
and does not follow from an affine reparametrization of the null generators.
Its necessity follows from the intrinsic Carrollian worldsheet geometry:
the relative Carroll--Weyl rescaling changes only the representative of the
same geometric structure and must therefore be gauged. Its admissibility is
a separate dynamical statement. At finite tension, the constraint algebra
contains an obstruction proportional to string tension squared, cf. \eqref{delta-C-T}, whereas this obstruction
vanishes in the tensionless limit and $C_3$ becomes first class. This
separates clearly why the constraint must be imposed from why it can be
imposed consistently.

The distinction is visible directly in the reduced phase space. The affine
ILST system with the two first-class constraints $C_1$ and $C_2$ has
$2(D-2)$ local phase-space degrees of freedom. The gauge-complete system
contains the additional first-class constraint $C_3=P\cdot X$ and its
associated gauge quotient. It therefore has $2(D-3)$ local phase-space
degrees of freedom, or $D-3$ local configuration-space degrees of freedom,
assuming a regular constraint surface and treating global zero modes
separately
\cite{Sheikh-Jabbari:2026cnj,Sheikh-Jabbari:2026tpf,
Sheikh-Jabbari:2026vqh}. Equivalently, the additional Carroll--Weyl
identification removes one local target-space degree of freedom before the
remaining ILST constraints and gauge identifications are imposed.

We close with three comments. First, the present analysis is restricted to
a classical null string in a flat Minkowski target space. Relations between
null strings and black hole or cosmological horizons have been investigated
in
\cite{Bagchi:2023cfp,Bagchi:2024rje,Bagchi:2026qpi}.
Extending the Carroll--Weyl gauging employed here to a generic curved target
space requires additional geometric input and is not addressed in the
present work.

Second, our discussion concerns classical null strings. Initial steps
toward their quantization have been taken in
\cite{Rasulian:2026jvg,Duary:2026rlo,Duary:2026lmk,Chen:2026cau}.
In particular, a discrete spectrum despite the absence of tension has been
reported in \cite{Rasulian:2026jvg}. The quantization of the complete
three-constraint system and the role of the Carroll--Weyl gauge symmetry at
the quantum level require further analysis.

Finally, the  ILST action is invariant under rigid target-space
translations,
\begin{equation}
 X^\mu(\tau,\sigma)\longmapsto X^\mu(\tau,\sigma)+a^\mu,
\end{equation}
whereas this symmetry is not manifest in the gauged action
\eqref{eq:main-CW-action}. The flat-space realization
$\delta_\chi X^\mu=\chi X^\mu$ is defined relative to a chosen target-space
origin. The status of translations in a more general realization of the
Carroll--Weyl gauging is beyond the scope of the present analysis.

\paragraph{Acknowledgments.}
We thank Alireza Akbari, Arjun Bagchi, Aritra Banerjee, Daniel Grumiller, Ulf Lindstr\"om, Ida Rasulian, Bo Sundberg and Shing Tung Yau for discussions. MMShJ acknowledges Iranian National Science Foundation (INSF) research chair grant No.~40451653. HY is supported in part by Beijing Natural Science Foundation under Grant No.~IS23013.

\appendix

\section{Intrinsic Carrollian geometry and conformal weights}
\label{app:geometry}

This appendix records the geometric calculation used in sections~\ref{sec:physical-motivation} and \ref{sec:CW-gauging}. Choose locally a Carrollian coframe $n_a,\ell_a$ and dual vectors $v^a,\ell^a$ satisfying
\begin{equation}
	v^an_a=1,
	\qquad v^a\ell_a=0,
	\qquad \ell^an_a=0,
	\qquad \ell^a\ell_a=1.
\end{equation}
The degenerate spatial metric and volume form are
\begin{equation}
	h_{ab}=\ell_a\ell_b,
	\qquad
	\varepsilon_{ab}=n_a\ell_b-n_b\ell_a,
	\qquad v^ah_{ab}=0.
\end{equation}
The same temporal and spatial line fields are represented after
\begin{align}
	n_a&\longmapsto e^{\chi_t}n_a,
	&v^a&\longmapsto e^{-\chi_t}v^a,
	\nonumber\\
	\ell_a&\longmapsto e^{\chi_s}\ell_a,
	&\ell^a&\longmapsto e^{-\chi_s}\ell^a.
\end{align}
The derived fields transform as
\begin{equation}
	h_{ab}\longmapsto e^{2\chi_s}h_{ab},
	\qquad
	\varepsilon_{ab}\longmapsto e^{\chi_t+\chi_s}\varepsilon_{ab}.
\end{equation}
Resolving the parameters into common and relative parts,
\begin{equation}
	\rho=\frac{\chi_t+\chi_s}{2},
	\qquad
	\chi=\frac{\chi_t-\chi_s}{2},
\end{equation}
shows that $\chi$ preserves the volume form while changing the relative normalization of the temporal and spatial directions.

Let $\epsilon^{ab}$ be the alternating symbol and define
\begin{equation}
	\mathfrak e=\epsilon^{ab}n_a\ell_b,
	\qquad
	\cV^a=|\mathfrak e|^{1/2}v^a.
\end{equation}
Then
\begin{equation}
	\mathfrak e\longmapsto e^{2\rho}\mathfrak e,
	\qquad
	\cV^a\longmapsto e^{-\chi}\cV^a.
\end{equation}
Thus $\cV^a$ is blind to the common rescaling and has weight $-1$ under the relative rescaling. With inert target coordinates,
\begin{equation}
	\delta_\chi S_{\rm ILST}
	=-\kappa\int_\Sigma\dd^2\sigma\,
	\chi\,\cV^a\cV^b\eta_{\mu\nu}\partial_aX^\mu\partial_bX^\nu,
\end{equation}
so the bare kinetic term does not descend to the equivalence class of Carrollian line-field representatives. In Minkowski space the simultaneous scaling $X^\mu\mapsto e^\chi X^\mu$ compensates the homogeneous variation, and a connection is required to localize it.

\section{Constraint algebra at zero tension}
\label{app:constraint-algebra}

For completeness, define the smeared constraints
\begin{equation}
	H[f]=\int\dd\sigma\,f\,\frac{P^2}{2},
	\qquad
	D[g]=\int\dd\sigma\,g\,P\cdot X',
	\qquad
	S[h]=\int\dd\sigma\,h\,P\cdot X,
\end{equation}
with $\{X^\mu(\sigma),P_\nu(\sigma')\}=\delta^\mu{}_{\nu}\delta(\sigma-\sigma')$. For a closed string, or boundary conditions that remove surface terms, their Poisson brackets are
\begin{subequations}
\begin{align}
	    \{H[h],H[k]\}&=0,\\
        \{D[f],D[g]\}&=D[fg'-gf'],\\
	\{H[f],D[g]\}&=H[fg'-gf'],\\
	\{D[g],S[h]\}&=S[gh'],\\
	\{S[h],H[f]\}&=2H[hf],\\
	\{S[h],S[k]\}&=0.
\end{align}
\end{subequations}
Thus $C_1$, $C_2$, and $C_3$ form a first-class system. Importantly, $C_1$ and $C_2$ also close to BMS$_3$ algebra by themselves. The algebra therefore establishes that $C_3$ is admissible at zero tension; its necessity follows separately from the requirement to gauge the complete intrinsic Carrollian representative redundancy.

At nonzero tension, the Hamiltonian constraint is $H_T=\tfrac12(P^2+T^2X'^2)$. Under the local scaling generated by $S[h]$, one finds on $H_T\approx0$
\begin{equation}
	\delta_hH_T\approx T^2\left(2hX'^2+h'X\cdot X'\right),
\end{equation}
which is not a combination of the usual tensile constraints for arbitrary $h$. This is the tensile obstruction that disappears in the singular $T\to0$ limit.
\section{Gauging the background conformal group?}
\label{app:gauging-CKV}

We discussed that the  gauged action \eqref{eq:main-CW-action} is the action with the Carroll-Weyl scaling of the worldsheet as its gauge symmetry. On the embedding coordinates $X^\mu$ the Carroll-Weyl scaling acts as a scaling, 
$X^\mu\to e^{\chi} X^\mu$. One may try to generalize this construction as follows. The ``conformal gauged action''
\begin{equation}
 S_{\rm K}
 =\frac{\kappa}{2}\int_\Sigma\dd^2\sigma\,
 \cV^a\cV^b\eta_{\mu\nu}D_aX^\mu D_bX^\nu,
 \label{eq:main-KCW-action}
\end{equation}
with
\begin{equation}
 D_aX^\mu=\partial_aX^\mu+\cW_a K^\mu,
 \label{eq:main-CW-K-derivative}
\end{equation}
which is invariant under the Conformal-Carroll--Weyl gauge transformations,
\begin{equation}\label{eq:main-K-CW}
 \delta_\chi\cW_a=-\partial_a\chi,
 \qquad
 \delta_\chi X^\mu=\chi K^\mu,
 \qquad
 \delta_\chi\cV^a=-\frac1D \chi \partial_\mu K^\mu\cV^a,
\end{equation}
if $K^\mu$ is satisfying the conformal Killing equation of the $D$ dimensional Minkowski space:
\begin{equation}
    \partial_\mu K_\nu+\partial_\nu K_\mu=\frac{2}{D} (\partial_\alpha K^\alpha) \eta_{\mu\nu}.
\end{equation}

For generic $D$ dimensional case,  $K^\mu$ which fall into three categories: (1) Isometries for which $\partial_\alpha K^\alpha=0$, these include translations $K^\mu= a^\mu=$const. and Lorentz transformations; (2) Dilatation, $K^\mu=X^\mu$. for which $\frac1D\partial_\alpha K^\alpha=1$; (3) Special conformal transformations for which $\frac1D \partial_\alpha K^\alpha= b_\mu X^\mu$ for constant vectors $b_\mu$. As \eqref{eq:main-K-CW} shows, for the isometries $\delta \cV^a=0$ and hence it is not gauging Carroll-Weyl scaling and, for special conformal transformations $\delta \cV^a=- \chi (b_\mu X^\mu) \cV^a$. This is not again a  Carroll-Weyl scaling, because Carroll-Weyl scaling must only involve worldsheet fields and not string embedding coordinates $X^\mu$. So, the only option which remains is $\frac1D \partial_\alpha K^\alpha=1$, which is precisely what we have considered. 

\end{document}